\documentclass[aps,prl,twocolumn,preprintnumbers,amsmath,amssymb,superscriptaddress,floatfix]{revtex4}
\usepackage[dvips]{graphicx}
\usepackage{mathptmx}
\usepackage{verbatim}

\begin{document}
\title{Carrier Density and Magnetism in Graphene Zigzag Nanoribbons}
\author{J. Jung} \email{jeil@physics.utexas.edu}
\affiliation{Department of Physics, University of Texas at Austin, USA}
\author{A. H. MacDonald}
\affiliation{Department of Physics, University of Texas at Austin, USA}
\date{\today{}}

\begin{abstract}
The influence of carrier density on magnetism in a zigzag graphene nanoribbon is studied in a 
$\pi$-orbital Hubbard-model mean-field approximation.
Departures from half-filling alter the magnetism, leading to 
states with charge density variation across the ribbon and 
parallel spin-alignment on opposite edges.  Finite carrier densities cause the 
spin-density near the edges to decrease steadily, leading eventually to the absence of magnetism.  
At low doping densities the system shows a tendency to multiferroic order 
in which edge charges and spins are simultaneously polarized.
\end{abstract}

\maketitle

\noindent 
\section{Introduction}
Graphene sheets and related carbon based nanomaterials 
have attracted attention recently after seminal experiments \cite{novoselov,pkim} 
revealed novel physics related to their unique electronic structure\cite{graphenereviews}.
In graphene nanoribbons\cite{fujita,nakada, waka, hikihara, dutta, leehosik, ezawa, brey, sasaki, son_gap, son_half, pisani,
 yazyev, joaquin,pohang,superexchange,han,chem_ribbon,etching,zarea1,zarea2}
lateral confinement leads to size quantization and to one-dimensional conduction channels
whose properties depend qualitatively on edge termination character.
Neutral zigzag terminated ribbons have attracted particular attention 
because they have a flat band, perflectly flat  in simple $\pi$-band models, pinned to 
the Fermi level. 
In self-consistent field (SCF) theories, including
{\em ab initio} spin-density-functional theories, the flat band leads to robust magnetic order.  
Ferromagnetic alignment of spins at the zigzag edges is predicted also in 
treatments going beyond mean field \cite{hikihara,dutta}.
Although the reliability of SCF theories is uncertain and  
not yet tested experimentally, interest in zigzag edge magnetism has remained strong because of
potential for interesting applications in nano-electronics 
\cite{pohang}.

Most  studies of the electronic structure of zigzag terminated graphene 
ribbons have focused on properties of the neutral system or 
systems with substitutional doping \cite{subdop}.
We study the role of gate voltage induced changes in carrier density, {\em i.e.} gate doping.
A related work in the low carrier doping regime with an additional neutralizing background 
charge explored the possibility of stable non-collinear magnetic states \cite{noncollinear}.
In neutral systems, SCF theories predict edge magnetization in 
graphene nanoribbons with opposite spin polarizations on opposite edges \cite{fujita, leehosik, pisani}.
In theoretical studies of locally gated zigzag ribbon junctions usually
the non-interacting electronic structure is assumed \cite{pn1,pn3,pn4,pn5,pn6,pn7,pn8},
neglecting the possibility of doping-dependent interaction-driven rearrangements.
In this work we show that gate doping leads to changes in charge distribution, 
spin configuration, and total net spin polarization, which are accompanied 
by important modifications in electronic structure.
\begin{figure}[h]
\begin{center}
\begin{tabular}{c}
\\
\\  \\
       \resizebox{80mm}{!}{\includegraphics[angle=90]{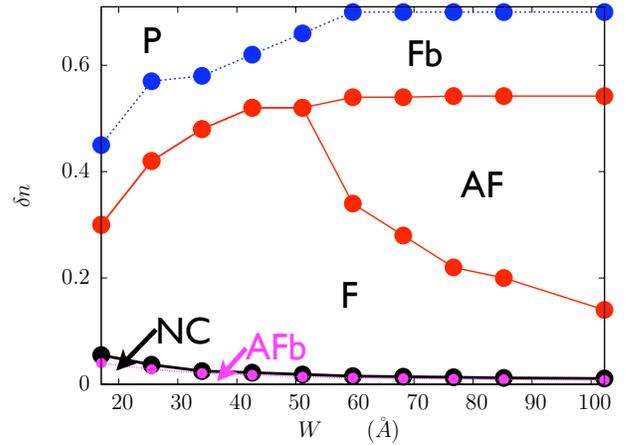}}  
\end{tabular}
   \caption{
(Color online)
Hubbard model SCF-theory phase diagram as a function of doping per length $\delta n$ (defined in the text)
and ribbon width $W$ for nearest-neighbor hopping  
$\gamma_0 = 2.6 eV$,
The on-site repulsion strength was chosen to 
have a value $U = 2 eV$ 
which reproduce the ribbon band gaps obtained
in the LDA-DFT calculations in reference [\onlinecite{joaquin}]. 
We used 1200 $k$-points for Brillouin-zone sampling. 
The energetic preference for opposite spins (AF) on opposite edges 
is replaced by a preference for parallel ($F$) spins at larger doping.
Above a critical doping $\delta n \sim 0.7$ the SCF calculation does not find magnetic states.
Solutions at finite doping sometimes (AFb and Fb) break the inversion symmetry of the ribbon. 
When non-collinear spin (NC) is allowed canted spin solutions midway between AF and F
configuration become energetically favored at low doping.
}
\label{fig:phase}
\end{center}
\end{figure}
Our study is based on the $\pi$-orbital Hubbard-model SCF theory 
for the magnetic properties of graphene nanostructures
\cite{yazyev1,joaquin,joaquin1,joaquin2,julio}, in which an electron
of spin $\sigma$ in site $i$ experiences a repulsive interaction
proportional to the density of opposite-spin electrons $n_{i\overline{\sigma}}$.
The Hubbard-model SCF theory is 
broadly consistent with DFT calculations when the interaction
parameter $U$ is chosen appropriately.
$\pi$-orbital Hartree-Fock theory reduces to the Hubbard model when
only the on-site Coulomb interactions are retained.
We have chosen to use a Hubbard interaction parameter $U = 2 eV$ 
which reproduces in the undoped case the band-gaps obtained by microscopic
density functional theory in the local density approximation.
This value is smaller than other estimates \cite{yazyev1}, but 
has been adopted with a similar motivation in some other recent work\cite{joaquin1,joaquin2}.

The Hubbard model mean-field Hamiltonian for each spin $\sigma$ is 
\begin{eqnarray}
H_{\sigma} = - \gamma_0 \sum_{\left< i,j \right>}  c^{\dagger}_{i \sigma}c_{j \sigma} +
U \sum_{i}  n_{i \overline{\sigma}} \, n_{i \sigma}   c^{\dagger}_{i \sigma} c_{i \sigma} 
+v_{ext} \sum_{i}   c^{\dagger}_{i \sigma} c_{i \sigma}  \nonumber
\end{eqnarray}
consist of a nearest neighbor tight-binding term with hopping $\gamma_{0} = 2.6 eV$
connecting lattice sites $i$ and $j$, the Hubbard term representing electron-electron
interactions, 
and an external potential term accounting for the interaction with the constant 
positive background charge proportional to a coefficient we choose to be $v_{ext} = -U$. 
Given the uncertainty of predictions implied by particular versions of SCF-theory,
the advantages of this relatively simple model often outweigh disadvantages.
Because the magnetism in zigzag ribbons is essentially one-dimensional,
we measure doping $\delta n$ in units of the number of excess electrons
per repeat distance $a = 2.46 \AA$ along the edge.
The corresponding areal density $\delta n_{2D} = \delta n / W$ where the ribbon width 
$W = \sqrt{3} N a/2$ and $N$ is the number of atom pairs 
per ribbon unit cell. 
\begin{figure}[tb]
\begin{center}
\begin{tabular}{c}
\resizebox{80.0mm}{!}{\includegraphics{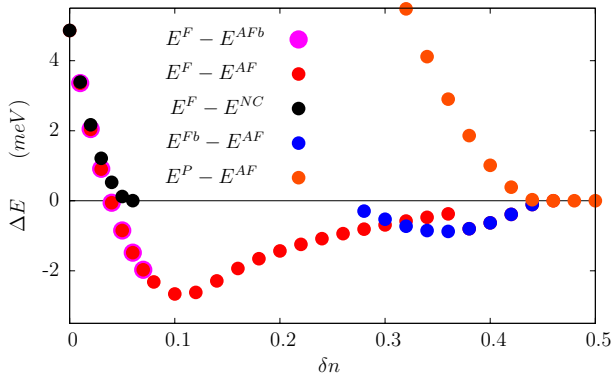}}
    \end{tabular}
\caption{(Color online)
Total energy differences per edge atom between the AF (or AFb) and F, Fb, P states as a function of doping $\delta n$
for a relatively narrow ribbon with $N=8$ atom pairs per unit cell. 
For low doping AFb type solutions with broken charge symmetry are energetically favored
over AF solutions although the energy difference is very small. The non-collinear (NC) spin 
solutions are lowest in energy in the weakly doped regime.
}
\label{fig:energy}
\end{center}
\end{figure}
\noindent 
\section{SCF solutions at finite doping}
The main players in zigzag edge magnetism are the flat band states 
which occupy one-third of the one-dimensional ribbon Brillouin-zone (BZ) and 
are localized\cite{brey} near the ribbon edges, most strongly so
near the BZ boundary $\left| k \right| \sim \pi /a$.
In the undoped SCF ground state, electrons of opposite-spin 
are localized near opposite edges of the ribbon and a gap\cite{superexchange} $\Delta \propto W^{-1}$ 
separates occupied valence and empty conduction band ribbon states.  By appealing to particle-hole symmetry 
we can limit our discussion of doping to the $n$-type case in which 
electrons start filling the conduction band.
Doping causes charge-density variation across the ribbon 
and to a complicated competition between band and interaction energies 
manifested by the variety of SCF equation solutions classified below. 
We label solutions as AF (opposite) or F (parallel) to indicate the relative alignment of spins on 
opposite edges.  
The label NC is used indicate non-collinear spin solutions.
The label b is applied for solutions 
which break inversion symmetry across the ribbon in a way which will be explained in more detail 
later. Finally we use the letter P to designate
a paramagnetic state with no local spin-polarization.  
The phase diagram in Fig. (~\ref{fig:phase}) illustrates the 
sequence of transitions AF $\rightarrow$ NC $\rightarrow$ F $\rightarrow$ Fb $\rightarrow$ P 
in narrower ribbons.  
In wider ribbons we find an additional AF state region between the 
F and Fb regimes.

\begin{figure}[tb]
\begin{center}
\begin{tabular}{c}
      \resizebox{80mm}{!}{\includegraphics[angle=90]{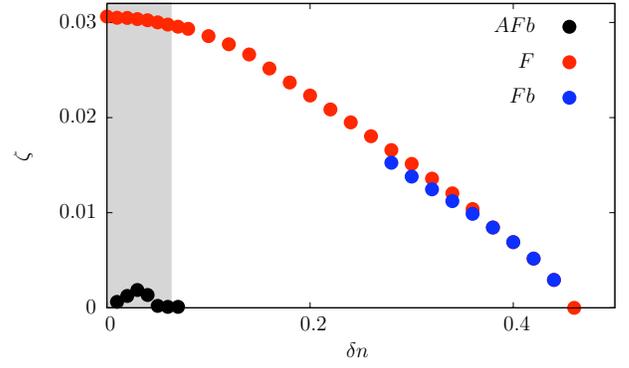}}  
    \end{tabular}
\caption{(Color online)
Net spin polarization  obtained from the total electron spin densities 
$\zeta =   \left( n_{\uparrow} -  n_{\downarrow}\right) / \left(  n_{\uparrow} +  n_{\downarrow} \right)$
for AFb, F and Fb solutions as a function of doping.
AFb solutions collapse into AF solutions with zero net spin polarization
for high enough doping. F and Fb configurations also progressively lose net spin polarization
as they approach the non-magnetic $P$ limit.
The shaded region represents the doping regime where non-collinear solutions are favored energetically.
}
\label{fig:spol}
\end{center}
\end{figure}

The total energy per unit cell consist of a sum over all the occupied single-particle
eigenvalues $\epsilon_{k \, m \, \sigma}$ labeled with $k$ and $m$ the band index
divided by $N_{K}$ the total number of $k$-points
minus a term to account for the double counting correction in the interaction
\begin{eqnarray}
\nonumber
E = \frac{1}{N_{K}}\sum_{k \, m \, \sigma}^{occ} \epsilon_{k \, m \, \sigma} - \frac{U}{2} \sum_{i \sigma}  n^{local}_{i \sigma} n^{local}_{i \overline{\sigma}}
\end{eqnarray}
where the occupations $n^{local}_{i \sigma}$ are evaluated in the local frame 
at lattice site $i$ where spin is diagonal. 
Their differences between different self-consistent solutions
are shown in Fig. (~\ref{fig:energy}) for a particular ($N=8$) ribbon 
width when only collinear spin solutions are considered. 
In the collinear scheme the energy associated with breaking inversion symmetry across the ribbons is 
always small and the main trend is a crossover from antiferromagnetic solutions at small $\delta n$ to 
ferromagnetic solutions for $\delta n \gtrsim 0.04$ to non-magnetic solutions for $\delta n \gtrsim 0.4$.
For the F type solutions and those with broken charge symmetry the system has a nonzero
net spin polarization as a function of doping density.  The doping dependence of 
spin-polarization is illustrated for the same $N=8$ ribbon 
width in Fig. (~\ref{fig:spol}).
\begin{figure*}[htb]
\begin{center}
\begin{tabular}{ccccc}
\resizebox{44.95mm}{!}{\includegraphics{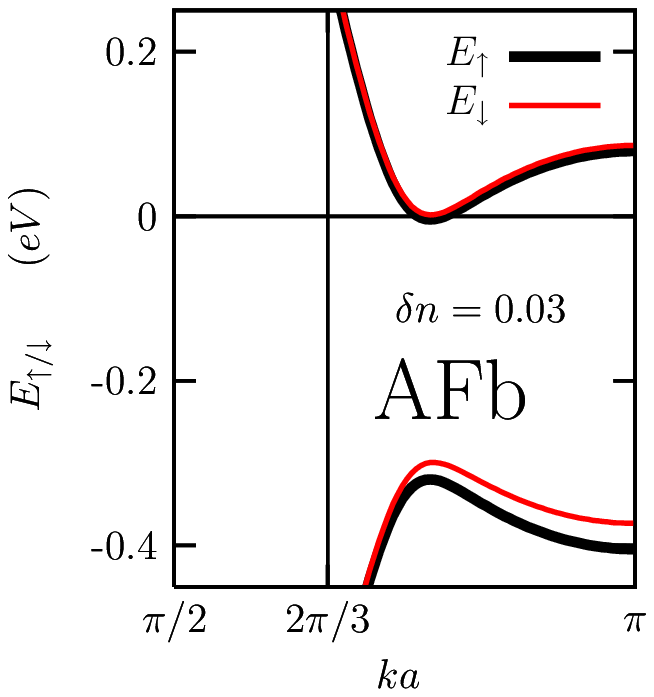}} &  \resizebox{32mm}{!}{\includegraphics{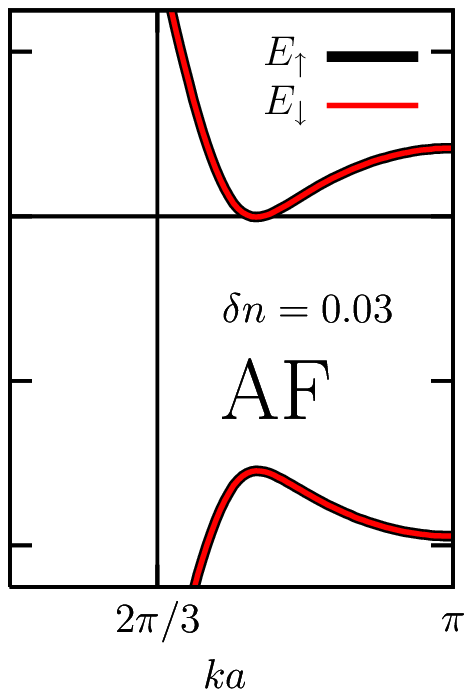}} &
 \resizebox{32mm}{!}{\includegraphics{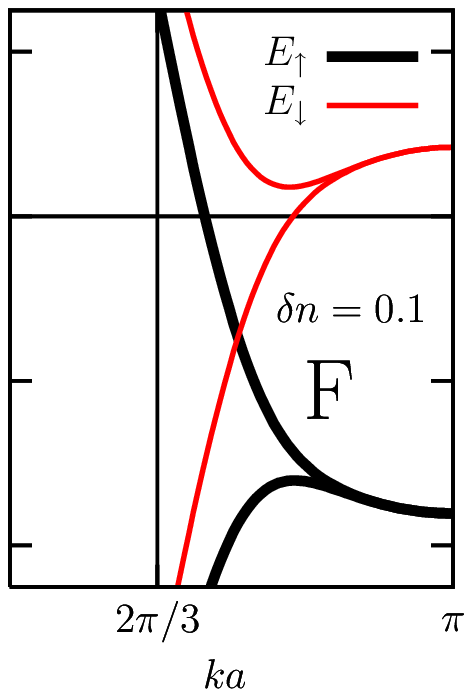}} & \resizebox{32mm}{!}{\includegraphics{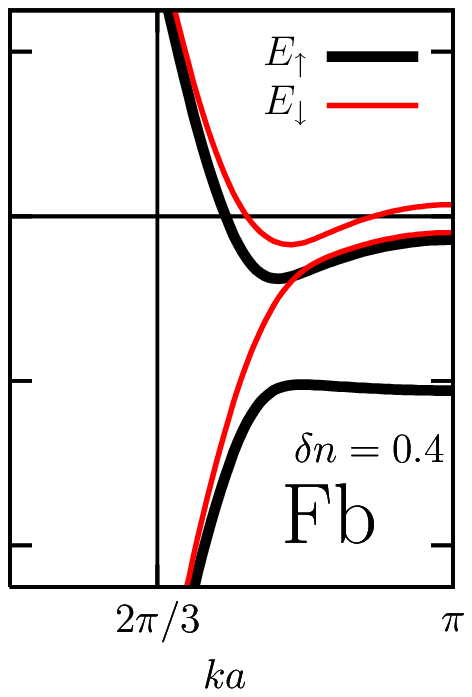}}
 & \resizebox{32mm}{!}{\includegraphics{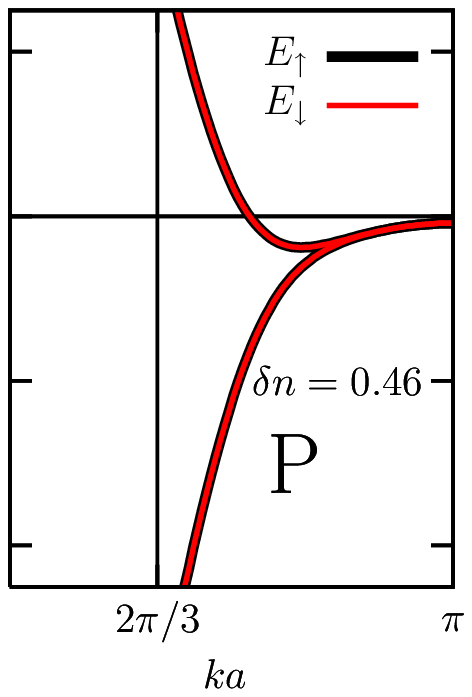}}   \\
 \\
\resizebox{44.7mm}{!}{\includegraphics{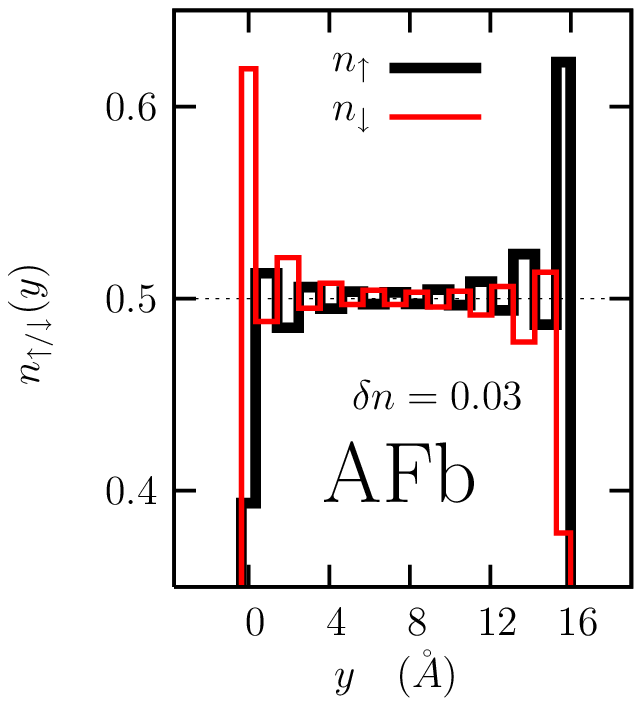}} &  \resizebox{32mm}{!}{\includegraphics{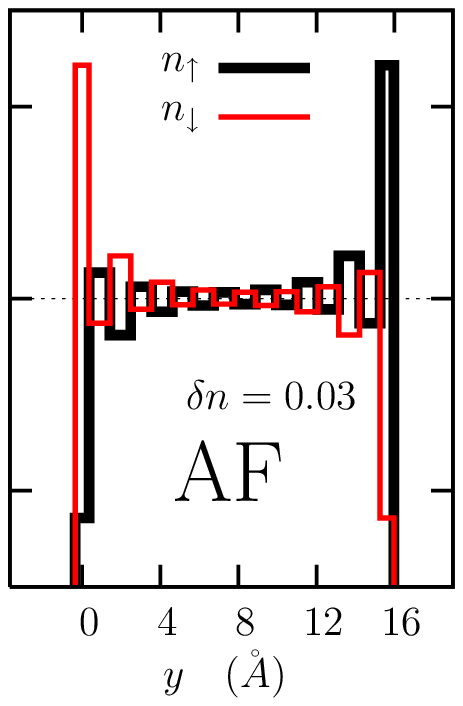}} &
  \resizebox{32mm}{!}{\includegraphics{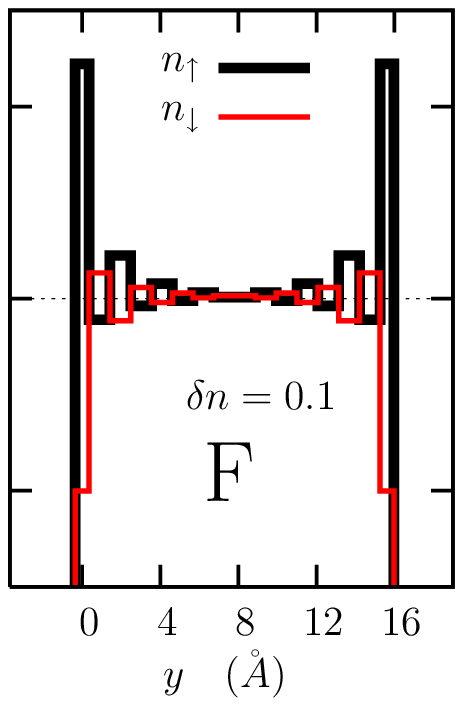}} & \resizebox{32mm}{!}{\includegraphics{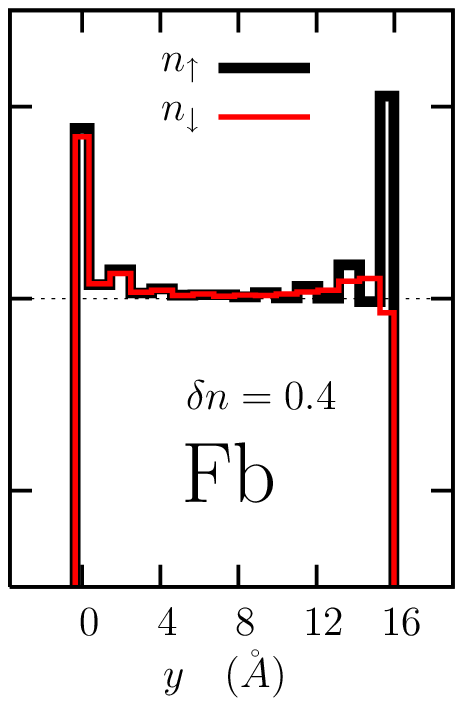}} &
  \resizebox{32mm}{!}{\includegraphics{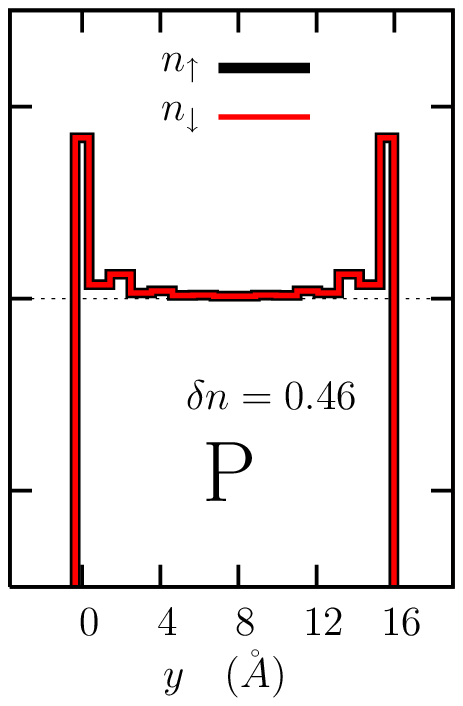}} \\
    \end{tabular}
\caption{(Color online)
{\em Upper row.} Band structures corresponding to AFb, AF, F, Fb and P spin collinear solutions of the 
Hubbard-model SCF equations for a zigzag nanoribbon with $N=8$ atom pairs in the unit cell.  
At finite doping the energy gain due to 
the gap present in the the AFb and AF solutions is reduced, favoring the F solution 
which does not have a gap.
{\em Lower row.}
Up and down spin electron occupation per lattice site in the unit cell across the ribbon.
The AFb configuration has broken charge distribution symmetry 
relative to the ribbon center due to an unequal occupation of up and down spin bands.
All mean-field bands are invariant under $k \to - k$.  We show only the portion of the 
1D BZ with states close to the Fermi level.
}
\label{fig:bands}
\end{center}
\end{figure*}
Each of the solution types identified in Fig.~(\ref{fig:phase}) is associated with particular 
electronic structure features which are illustrated in Fig.~(\ref{fig:bands}).
For the AF solution, finite doping requires that states above 
the interaction induced gap be occupied.  For small doping 
electrons start occupying  states near the conduction band minima.
(See Fig. (\ref{fig:bands}).)
These additional electrons suffer a large energy penalty due to the 
neutral solution band-gap and have lower energy 
when spin-polarized.
The resulting half metallic solution in the spin-collinear scheme 
implies a non-zero overall spin polarization in the system and is accompanied by a 
breaking of charge distribution symmetry around the ribbon center.
This asymmetric charge distribution is a combined effect of the 
net spin polarization and the character of the AF solution at the 
neutrality point, in which electrons with opposite spin polarizations 
are concentrated on opposite edges \cite{superexchange}. 
If both up and down spin bands were equally occupied there would 
be no charge distribution asymmetry around the ribbon center.
 
In the low doping regime a non-collinear spin-order  that continuously bridges 
the intermediate situation between the neutral AF configuration and mostly F 
configuration at higher doping is \cite{noncollinear} a possibility.  
In the version of the Hubbard model mean-field theory which allows for 
non-collinear spin denisties we must allow for the possibility that 
the average spin polarization on different lattice sites points in different directions \cite{haney}. This allows a larger
variational space within a single Slater determinant approximation and can potentially lead to 
lower energy solutions, but
the spin label becomes undefined for each single-particle wave function.
We verified that non-collinear spin solutions are favored energetically \cite{noncollinear} in the
Hubbard model calculations for low doping region 
and that the transition to F configuration happens at doping densities typically about 
$20 \%$ higher than when only collinear solutions are considered. 
The angle between the spin densities on opposite edges and
the band structure of the non-collinear state are represented in Fig. 5.
\begin{figure}[htb]
\begin{center}
   \begin{tabular}{c}
          \resizebox{80mm}{!}{\includegraphics[angle=90]{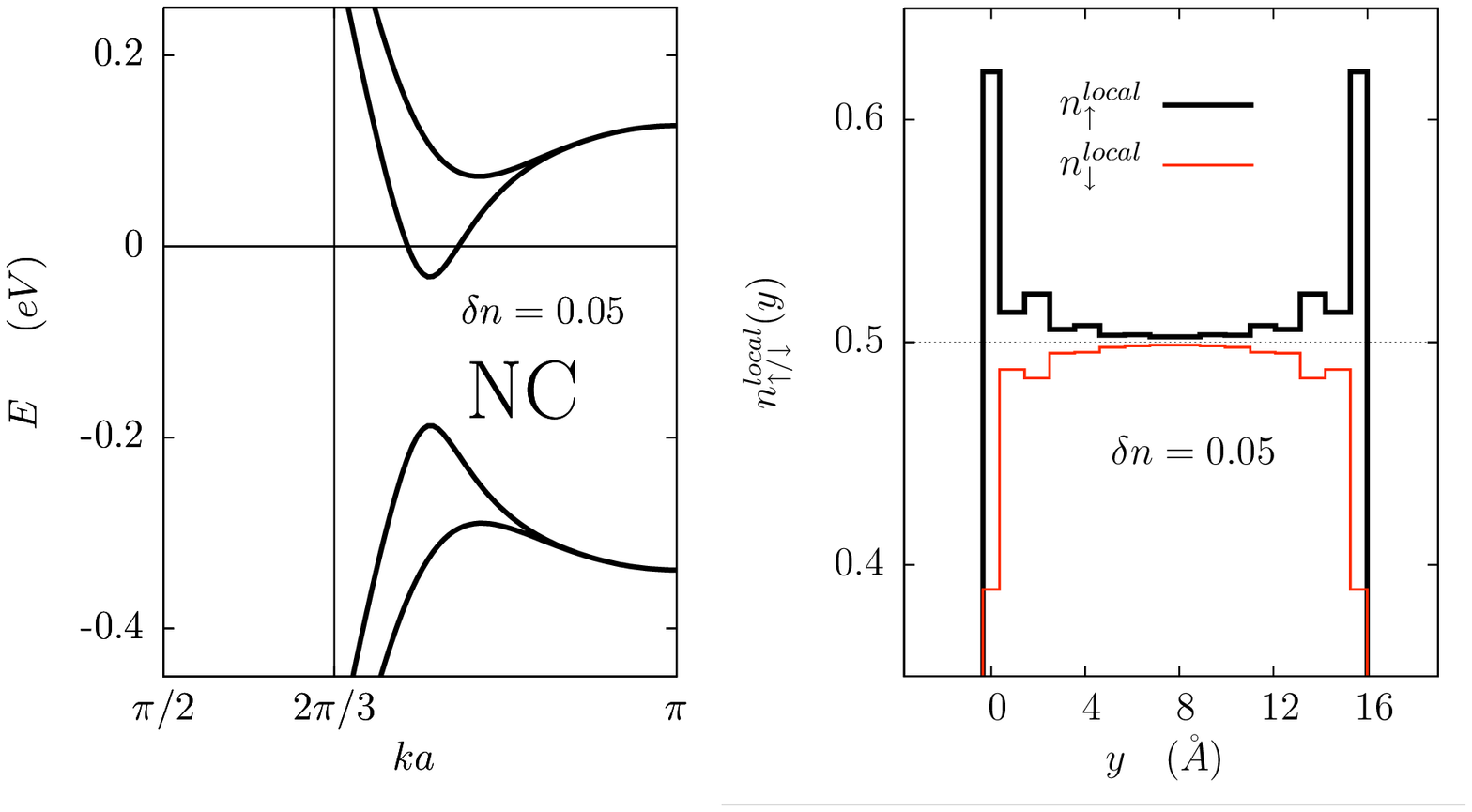}}  
          \\ \\
       \\  \\
       \resizebox{80mm}{!}{\includegraphics[angle=90]{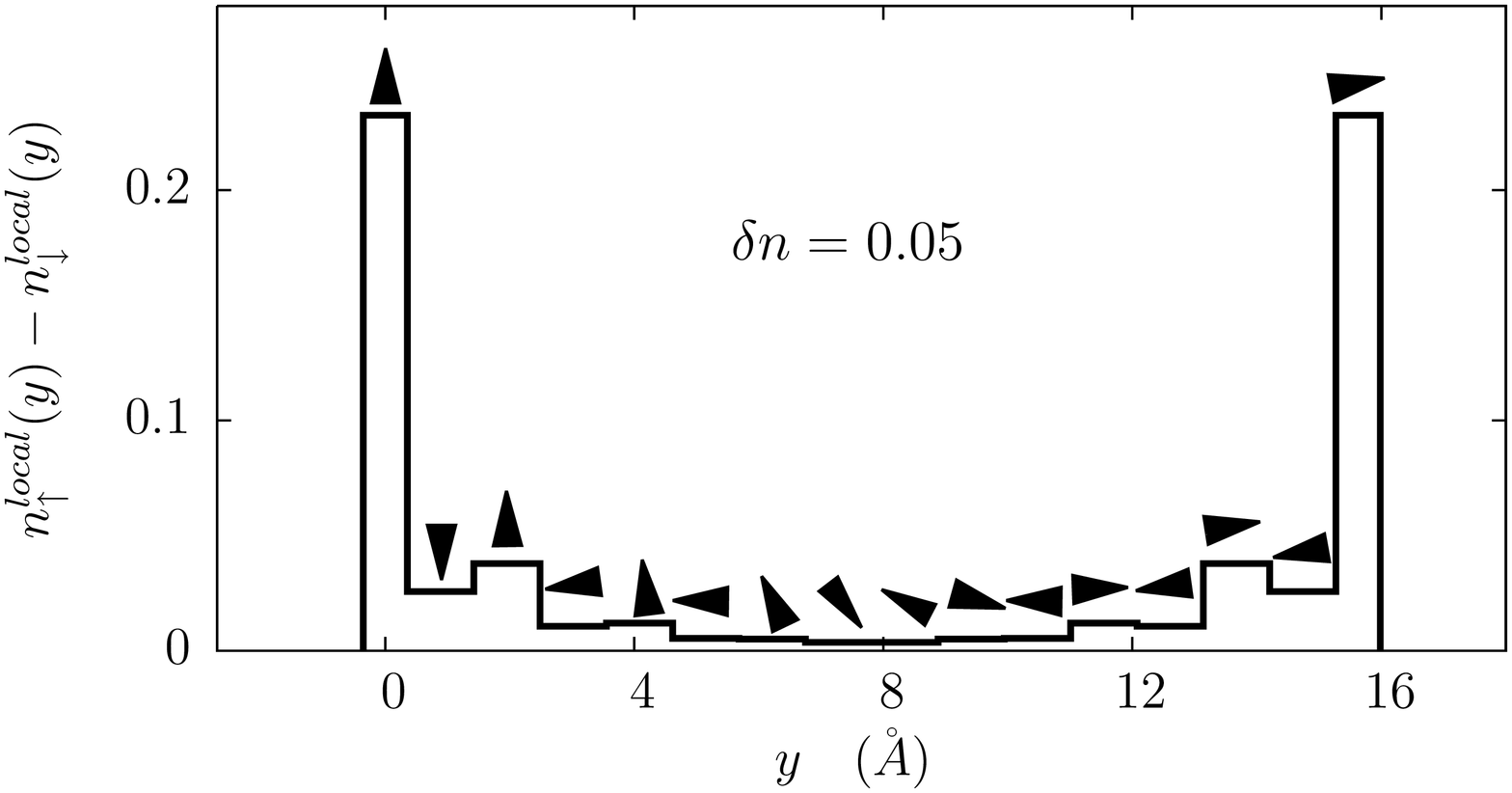}}  
    \end{tabular}
\caption{(Color online)
In the weakly doped region canted spin orientations develop 
in order to minimize the total energy when non-collinear solutions are allowed.
{\em Upper row.} Band structure and spin resolved electron occupation
per lattice for a zigzag ribbon with $N=8$ and $\delta n = 0.05$. 
The occupation and spin polarization at each lattice site
are represented in a local frame where the spin is diagonal.
{\em Lower row.}
Spin polarization and relative
orientation of the spin direction between different lattice sites
represented with the arrow heads.
}
\label{fig:bands_nc}
\end{center}
\end{figure}

In the intermediate doping regime the total energy is minimized 
by solutions which are more similar to the F neutral-ribbon configuration\cite{superexchange}
which do not have an energy gap, and are therefore favored by doping.
This transition to F-like solutions
occurs already at a relatively small value of doping $\delta n \simeq 0.06$.
The states that are occupied first at finite doping are those near the valley 
points $\left| k \right| = 2\pi/3a$ 
that are\cite{superexchange} spread across the ribbon and 
therefore control the exchange coupling between opposite edges. 
The $W$-scaling rules of the energy bands
near the valley points \cite{superexchange} are consistent with the 
$W^{-1}$ decay law of the  threshold doping at which the transition
to F-like transition occurs in our numerical phase diagram. 
In electronic structures with dominantly F character the charge distribution symmetry 
around the ribbon center is preserved.
In this case every occupied states,
up or down spin and valence or conduction edge band, 
has a symmetric distribution of electron density around the ribbon center. 
When the doping is sufficiently large, however, we find a broken charge symmetry 
solution that we label as Fb.  In this state one of the occupied conduction bands
has AF (unbalanced across the ribbon) rather than F (balanced across the ribbon)
character. 
In addition to these solutions, we find that for wide ribbons 
there is an intermediate doping region in which AF solutions have lower total energy than 
the F solutions before the Fb solution is stabilized.  The 
difference in energy between different magnetic solutions is small at
intermediate and large doping. 

In the high doping regime the magnetic features of the system 
progressively disappear as the edge state bands become filled.
The ribbon is found to turn paramagnetic above a critical value 
that increases with the ribbon width and saturates 
around $\delta n_c \sim 0.7$. 
Considering that edge localized states in the conduction bands
with $k$-points near $2\pi/3a \leq \left|k\right| \leq \pi/a$ span approximately
$1/3$ of the whole Brillouin zone we find that 
the total amount of doping electrons required to fill completely the edge 
for both up and down spins is $2/3$, an amount that can be surpassed near the
mentioned doping saturation limit.

\noindent  
\section{Discussion}
The AF state of zigzag nanoribbons has the unusual feature that 
inversion symmetry across the ribbon is broken in opposite senses in the two spin sub-systems 
\cite{joaquin,superexchange}.
Our calculation suggest that in low doping regime the system can easily develop solutions 
with a charge density that is distributed asymmetrically across the ribbon,
creating an interesting and unusually strong type of 
multiferroic behavior \cite{multiferroic,joaquin} in which spin polarization and charge density are coupled.
We expect that transport properties can correspondingly be manipulated in 
interesting interrelated ways by both external magnetic fields and external electric fields 
directed across the ribbon.  Edge transport should be strongly suppressed,
for example, when a transverse electric field is applied which 
has opposite orientations on opposite ends of a ribbon.

Above a certain critical doping density, 
which is inversely proportional to the ribbon width $W^{-1}$, 
we find that the system undergoes a 
transition to a F configuration in which opposite edges have parallel spin polarizations. 
When doping is increased further the spin-configuration is altered yet again,
restoring inversion symmetry breaking across the ribbon.  In this 
high-doping regime the total magnetic condensation energy is small and 
the energy differences between different magnetic configurations is small.
Eventually at sufficiently high doping the Hubbard-model SCF equations 
have only paramagnetic solutions.  

The $Si O_2$ substrates on which exfoliated graphene samples are usually 
prepared have electron density inhomogeneities  \cite{yacobi} of the order of 
$n_{fluc} \sim 10^{11} cm^{-2}$  
that can extend over lengths of the  order of $L \simeq 1 \mu m$.
In the limiting case of ribbons with this same width as the puddle sizes
a rough estimate of doping per unit lattice constant $a$ within each
puddle can be evaluated with the product of these two quantities
\begin{eqnarray}
\delta n_{fluc} \sim  L \cdot n_{fluc} \sim 0.25 / a   \nonumber
\end{eqnarray}
This amount of doping can influence the spin configurations in the system and the presence 
of these random perturbations is therefore expected to appreciably weaken the tendency towards 
magnetic order, especially for wide ribbons. 
For this reason we should expect a better chances of detecting edge magnetism in suspended ribbons 
which have much weaker electron density fluctuations.
 
The long-ranged character of the Coulomb interaction, neglected in the present work,
is expected to introduce important changes in the details of electronic structure 
especially in the regions in which states with different 
charge and spin configurations compete closely.  The discrepancies can be more acute than 
in the neutral case because the inadequacy of short ranged screening can be 
more relevant when the occupation of each lattice site in the unit cell becomes 
inhomogeneous as we depart from half filling.
Nevertheless, it is also likely that several qualitative features of the solutions 
are still correctly captured by the Hubbard model and therefore can 
provide useful hints on the actual behavior of the magnetic configurations 
in ribbons as a function of doping.
Even though fluctuation effects we have neglected may work against the formation of long-range
order in these 1D-magnets, the unusually stiff ferromagnetic alignment of the spins predicted 
by mean field theories for zigzag ribbons \cite{yazyev} suggests 
that magnetic order could be possible and should be manifested in some way in experiments. 
  
{\em Acknowledgments.}  
We gratefully acknowledge helpful discussions with P. M. Haney,  T. Pereg-Barnea,
K. T. Delaney, and P. Rinke.
Financial support was received from the Welch Foundation, NRI-SWAN, ARO, DOE
and the Spanish Ministry of Education through the MEC-Fulbright program.

\end{document}